\documentclass[journal]{IEEEtran}

\usepackage{cite}
\usepackage{amsmath,amssymb,amsfonts}
\usepackage{mathtools}
\usepackage{graphicx}
\usepackage[caption=false,font=footnotesize]{subfig}
\usepackage{url}
\usepackage{placeins}
\usepackage{xcolor}

\graphicspath{{figure/}}
\DeclareGraphicsExtensions{.pdf}

\newcommand{\vctr}[1]{\mathbf{#1}}
\newcommand{\norm}[1]{\left\lVert#1\right\rVert}
\newcommand{\tauE}{\tau_E}

\newif\ifshowrevisions
\showrevisionstrue

\begin{document}

\title{Accelerated S-NFC for Million-Chaff RCS Computation Using Low-Rank Compression of Concatenated Block Rows}

\author{Dong-Yeop~Na,~\IEEEmembership{Member,~IEEE}, Somyeong~Lee,~\IEEEmembership{Student~Member,~IEEE}, and
Chung~Hyun~Lee,~\IEEEmembership{Member,~IEEE}%
\thanks{

Somyeong~Lee and Chung~Hyun~Lee are with the Division of Semiconductor and Electronics Engineering, Hankuk University of Foreign Studies, Yongin, Gyeonggi-Do, 17035, South Korea. (e-mail: 4350579@hufs.ac.kr, chunghyun.lee@hufs.ac.kr)

Dong-Yeop Na is with the Department of Electrical Engineering, Pohang University of Science and Technology, Pohang, Gyeongsangbuk-do 37673, South Korea (e-mail: dyna22@postech.ac.kr)
}}

\markboth{Preprint}%
{Na \MakeLowercase{\textit{et al.}}: Accelerated S-NFC for Million-Chaff RCS Computation}

\maketitle

{
\begin{abstract}
Sparsification via neglecting far-field coupling (S-NFC) enables fast full-wave radar-cross-section analysis of large-scale chaff clouds by retaining only significant local electromagnetic interactions. 
This letter further accelerates S-NFC by concatenating the retained off-diagonal interaction blocks associated with each receiving chaff element and applying a joint low-rank factorization with a shared receiving-side basis. 
Exact self interactions are preserved, while repeated chaff templates reuse precomputed lower--upper factorizations of the self-interaction blocks. 
The compressed formulation reduces retained-coupling storage and matrix--vector multiplication cost and also decreases the number of iterations required by the generalized conjugate residual solver. 
Numerical tests with 100,000 chaff elements demonstrate sub-$1\%$ complex-far-field error for low-rank approximations in sparse regimes and identify a practical self-only limit at sufficiently large mean spacing. 
For a one-million-chaff plume, the proposed compressed S-NFC achieves a $6.92\times$ end-to-end speedup over uncompressed S-NFC while storing only $6.60\%$ of the retained coupling, with a complex-far-field error of $0.253\%$.
\end{abstract}


\IEEEpeerreviewmaketitle

\section{Introduction}

\IEEEPARstart{C}{haff-cloud} radar-cross-section (RCS) prediction is computationally challenging because practical clouds may comprise an enormous number of conducting elements with diverse orientations, spatial distributions, and lengths.
To enable large-scale analysis, various reduced-order and approximate approaches have been developed.
Equivalent-conductor and effective-medium formulations efficiently represent the collective response of chaff clouds and have been extended to arbitrary orientation distributions, dynamic scenarios, and real-time cloud modeling~\cite{Seo2010GEC,Seo2012Dynamic,Kim2023Access,Kim2023RealTime}.
Analytical/statistical treatments have characterized propagation and ensemble scattering of chaff clouds~\cite{Marcus2007,Marcus2015}, while discrete electromagnetic (EM) models enable wideband and bistatic analysis~\cite{Alvarez2020DCCM}.
Vector radiative-transfer approaches have further addressed large-scale multiple scattering in airflow-dependent and sea-surface environments~\cite{Zuo2020VRT,Zuo2021VRT,Zheng2023}.
Although computationally efficient, these approaches do not generally provide an element-resolved full-wave solution of the induced currents and EM interactions for a particular cloud realization.

Full-wave thin-wire approximation (TWA) formulations based on the electric-field integral equation (EFIE) and method of moments (MoM)~\cite{gibson2024method} directly resolve the induced currents and inter-element interactions, but their application to large chaff clouds is constrained by the computational and memory costs of the resulting densely coupled impedance system.
Previous studies have shown that interactions between sufficiently separated chaff elements become progressively weaker~\cite{garbacz1975,peebles1984bistatic,Marcus2015,Zuo2022Coupling,Lee2025AWPL}.
For sufficiently sparse clouds, previous studies have empirically observed weak long-range coupling at mean inter-element spacings on the order of $2\lambda$, although the effective coupling range may vary with the specific chaff configuration and EM conditions~\cite{Zuo2022Coupling,Lee2025AWPL}.
This coupling locality provides a physical basis for truncating weak long-range interactions while retaining significant local coupling.
Building on this property, sparsification via neglecting far-field coupling (S-NFC) was introduced to sparsify the MoM impedance matrix, substantially reducing the computational and memory burden while retaining the element-resolved full-wave formulation~\cite{Lee2025AWPL}.

Recent developments have further enabled multiscale RCS computation and real-time analysis of dynamically evolving million-scale chaff clouds~\cite{Song2026MCCRCS,Lee2026SciRep}.
Data-driven acceleration has also been demonstrated using a graph neural network surrogate of S-NFC~\cite{Zhao2026GNN}.
While offering substantial acceleration, such surrogate approaches rely on S-NFC-generated training data, motivating further optimization of the underlying physics-based solver itself.
Despite its sparsification, S-NFC retains off-diagonal coupling blocks that can impose substantial memory and iterative-solver costs at million-element scale.
Low-rank techniques, including adaptive cross approximation, are well established for MoM interaction matrices~\cite{Heldring2013}; however, conventional pairwise block compression does not explicitly exploit the repeated chaff structure and block-row organization of S-NFC.

In this work, we further accelerate S-NFC by introducing a receiving-chaff-wise low-rank compression of the retained coupling operator. 
Unlike conventional pairwise block compression, all retained off-diagonal interactions associated with a receiving chaff are concatenated and jointly factorized, enabling neighboring source interactions to share a common receiving-side basis. 
This formulation reduces both memory and matrix--vector multiplication costs while preserving exact self interactions, and can additionally accelerate iterative convergence. Extensive validations include 100,000-chaff density and morphology studies, straight/bent/twisted-bent template tests, and one-million-chaff uniform and plume clouds. 
Both fixed- and adaptive-rank strategies are evaluated using consistent complex-far-field error and end-to-end runtime metrics. 
The results demonstrate that very low ranks are sufficient in sparse regimes and that the achievable compression and speedup are governed by the local coupling graph and cloud morphology. 
The $e^{+j\omega t}$ time convention is used throughout the letter.

\section{S-NFC with Exact Reuse and Concatenated-Row Compression}

Consider a cloud of $N_c$ chaff elements with thin-wire current coefficients
\begin{equation}
\vctr x=[\vctr x_1^T,\ldots,\vctr x_{N_c}^T]^T,
\qquad \vctr x_i\in\mathbb C^{n_i}.
\label{eq:xblock}
\end{equation}
A Galerkin TWA-EFIE discretization gives $\vctr Z\vctr x=\vctr b$.
Let $\mathcal N_i$ denote the source elements retained by S-NFC for receiving element $i$.
Then
\begin{equation}
(\vctr D_s+\vctr O)\vctr x=\vctr b,
\quad
\vctr D_s=\operatorname{blockdiag}(\vctr Z_{11},\ldots,\vctr Z_{N_cN_c}),
\label{eq:snfc_system}
\end{equation}
where $\vctr O$ contains $\vctr Z_{ij}$ only for $j\in\mathcal N_i$.

The S-NFC sparsity pattern is determined geometrically before any low-rank approximation is applied. For cutoff distance $d_c$,
\begin{equation}
\mathcal N_i=\{j\neq i:d_{ij}\le d_c\},
\label{eq:neighbor_set}
\end{equation}
where $d_{ij}$ denotes the separation used by the S-NFC interaction criterion.
The same $d_c=3\lambda$ graph is used by the uncompressed and compressed systems; compression changes only the representation of the already retained coupling.

Repeated chaff templates have identical self blocks under rigid translation and rotation. With template index $s(i)$,
\begin{equation}
\vctr Z_{ii}=\vctr Z_{\rm self}^{(s(i))},\qquad
\vctr P_s\vctr Z_{\rm self}^{(s)}=\vctr L_s\vctr U_s,
\label{eq:selfreuse}
\end{equation}
so only one self block and its LU factors are stored for each distinct template.
The exact template solve is $\mathcal S_s(\vctr v)=\vctr U_s^{-1}\vctr L_s^{-1}\vctr P_s\vctr v$.

For receiving element $i$, the retained mutual-coupling blocks are concatenated as
\begin{equation}
\vctr O_i=
\begin{bmatrix}\vctr Z_{ij_1}&\cdots&\vctr Z_{ij_{m_i}}\end{bmatrix}
\in\mathbb C^{n_i\times q_i},
\quad q_i=\sum_{j\in\mathcal N_i}n_j.
\label{eq:concatrow}
\end{equation}
Because all blocks in $\vctr O_i$ share the same receiving element and testing functions, their column spaces exhibit redundancy across source elements, motivating a common receiving-side basis.
The prescribed-rank approximation is
\begin{equation}
\vctr O_i\approx\widehat{\vctr O}_i^{(r)}
=\widetilde{\vctr U}_i\vctr V_i^H,
\qquad r_i=\min(r,n_i,q_i),
\label{eq:rowsvd}
\end{equation}
where $\widetilde{\vctr U}_i$ absorbs the retained singular values; $r_i=0$ for empty rows.

For adaptive compression, let $\sigma_{i,1}\ge\sigma_{i,2}\ge\cdots$ be the singular values of $\vctr O_i$. The row rank is
\begin{equation}
r_i=\min\left\{r:
\frac{\sum_{k=r+1}^{p_i}\sigma_{i,k}^{2}}
     {\sum_{k=1}^{p_i}\sigma_{i,k}^{2}}
\le \tauE\right\},
\qquad p_i=\min(n_i,q_i).
\label{eq:adaptive_rank}
\end{equation}
Here $\tauE$ is the discarded singular-value-energy tolerance and is unrelated to the far-field error $\epsilon_E$ defined later.
We also use $r=0$ as a diagnostic block-diagonal-only limit:
\begin{equation}
\widehat{\vctr O}^{(0)}=\vctr 0,
\qquad \vctr x_0=\vctr D_s^{-1}\vctr b.
\label{eq:rank0}
\end{equation}
Thus, all mutual coupling is removed while the exact self-block solution is preserved; this case quantifies when mutual coupling itself becomes negligible for the selected observable.

The preconditioned system is
\begin{equation}
\left(\vctr I+\vctr D_s^{-1}\widehat{\vctr O}\right)\vctr x
=\overline{\vctr b},
\qquad \overline{\vctr b}_i=\mathcal S_{s(i)}(\vctr b_i),
\label{eq:preconditioned_system}
\end{equation}
and each compressed block-row action is
\begin{equation}
\vctr y_i=\vctr x_i+
\mathcal S_{s(i)}\!\left[\widetilde{\vctr U}_i
(\vctr V_i^H\vctr x_{\mathcal N_i})\right].
\label{eq:compressed_mvm}
\end{equation}
Thus, self interactions and the block-diagonal preconditioner remain exact; only retained mutual coupling is approximated.
For common block size $n$, rank $r$, and mean neighbor count $\bar m$, neglecting indexing and metadata overhead, the compressed-to-full retained-coupling storage ratio is approximately
\begin{equation}
\rho_r\simeq\frac{r(1+\bar m)}{n\bar m}.
\label{eq:complexity_ratio}
\end{equation}

\section{Numerical Results}

Each reference is the uncompressed S-NFC solution for the same realization, excitation, $3\lambda$ cutoff, and stopping criterion.
With complex bistatic far-field vectors $\vctr e_\theta$ and $\vctr e_\phi$, all quantitative accuracy data use
\begin{equation}
\epsilon_E=\left[
\frac{\norm{\widehat{\vctr e}_\theta-\vctr e_\theta^{\rm ref}}_2^2+
\norm{\widehat{\vctr e}_\phi-\vctr e_\phi^{\rm ref}}_2^2}
{\norm{\vctr e_\theta^{\rm ref}}_2^2+
\norm{\vctr e_\phi^{\rm ref}}_2^2}
\right]^{1/2}.
\label{eq:efield_error}
\end{equation}
The vectors in \eqref{eq:efield_error} stack the same 720 complex-field observation samples. Because the reference is uncompressed S-NFC, $\epsilon_E$ isolates low-rank compression or block-diagonal-only error; it is not the total S-NFC error relative to the original dense MoM system.

The density and morphology tests use $f=8.427$~GHz, straight $\lambda/2$ chaff elements, 20 segments (19 unknowns) per element, incidence $(\theta,\phi)=(90^\circ,0^\circ)$, and the cut $\theta_{\rm obs}=90^\circ$, $\phi_{\rm obs}=0^\circ:0.5^\circ:359.5^\circ$.
Restartedgeneralized conjugate residual (GCR) uses a relative tolerance of $10^{-4}$; production runs use 120 OpenMP threads on two Intel Xeon Gold 6530 processors and 256~GB memory.

For the density/morphology and one-million-chaff sweep experiments, the total runtime $T$ includes matrix assembly, 360 right-hand-side solves with GCR, 360-point monostatic evaluation, 720-point bistatic evaluation, preconditioning, and I/O; full and compressed runs use identical workloads. The reported speedup is $
S={T^{\rm full}}/{T^{\rm comp}}.$


\subsection{Density and Morphology Dependence}

Fig.~\ref{fig:density} summarizes the 100,000-chaff sweep versus measured mean nearest-neighbor spacing $\bar d_{\rm NN}$.
At $0.5\lambda$, fixed ranks 1, 2, and 4 give $\epsilon_E=2.64\%$, $1.86\%$, and $0.951\%$, respectively, while adaptive $\tauE=0.01$ gives $0.919\%$.
At $2\lambda$, rank one gives $0.187\%$; beyond approximately $6\lambda$, the compressed curves approach $1.3\times10^{-2}\%$.
For $r=0$, the errors are $27.17\%$, $1.15\%$, $0.602\%$, and $0.383\%$ at $\bar d_{\rm NN}/\lambda=0.5,2,3,$ and $4$, respectively. Thus, for this observable, the block-diagonal-only model is a practical self-only approximation in the tested dilute regime $\bar d_{\rm NN}\ge3\lambda$, where $\epsilon_E<1\%$, although this threshold is not universal.

\begin{figure}[!t]
\centering
\includegraphics[width=0.94\columnwidth]{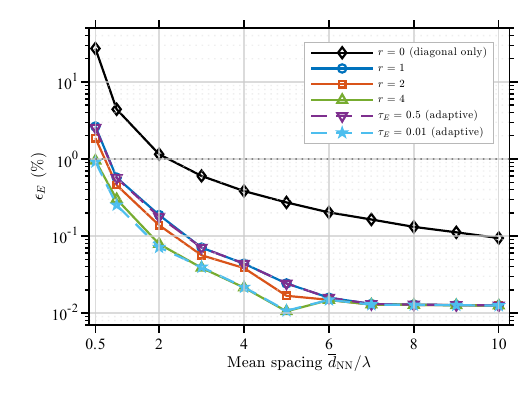}
\caption{Complex-far-field error versus measured mean nearest-neighbor spacing. Fixed ranks include the block-diagonal-only limit $r=0$; two representative adaptive tolerances are also shown.}
\label{fig:density}
\end{figure}

Fig.~\ref{fig:geometry} visualizes the two 100,000-chaff morphology cases at matched mean nearest-neighbor spacing $2\lambda$.
The uniform sphere has isotropic orientations, whereas the falling plume is spatially elongated and strongly biased toward horizontal chaff axes.
The mean/95th-percentile neighbor counts within $3\lambda$ are $2.41/5$ for the uniform sphere and $5.67/16$ for the plume, showing that matched mean spacing does not imply a matched local coupling graph.

\begin{figure}[!t]
\centering
\subfloat[Uniform sphere]{\includegraphics[width=0.9\columnwidth]{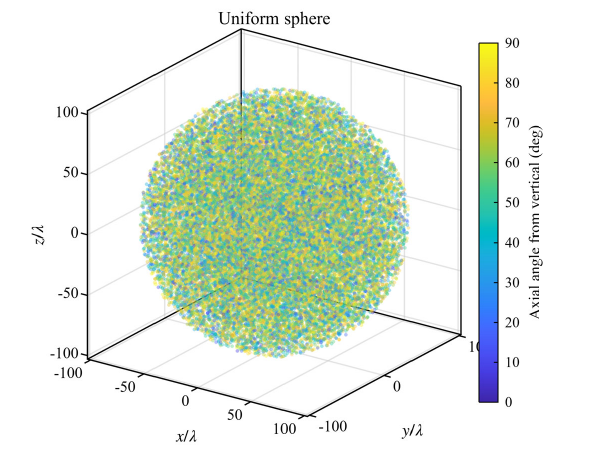}\label{fig:geometry_uniform}}\\[-2pt]
\subfloat[Falling plume]{\includegraphics[width=0.9\columnwidth]{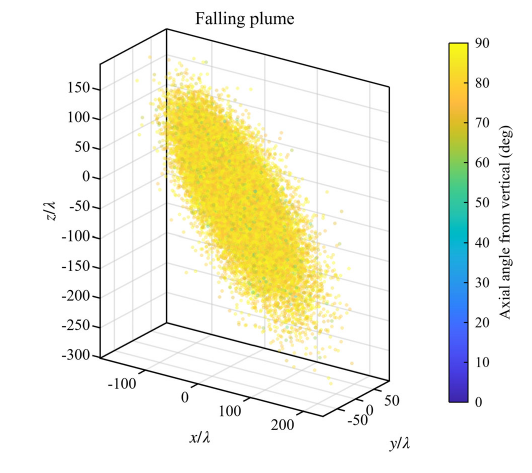}\label{fig:geometry_plume}}
\caption{Three-dimensional 100,000-chaff morphology at matched mean spacing $2\lambda$. Color denotes the acute axial angle from the vertical direction.}
\label{fig:geometry}
\end{figure}

Fig.~\ref{fig:ne1} compares fixed and adaptive compression for these clouds.
For fixed rank one, the uniform/plume errors are $0.187\%/0.845\%$, with end-to-end speedups $3.17\times/4.74\times$.
The block-diagonal-only errors are $1.153\%/3.686\%$; the larger plume error reflects its denser local coupling graph and greater sensitivity to mutual coupling.
For adaptive compression, tightening $\tauE$ decreases error while reducing speedup; the plume remains consistently more difficult than the uniform sphere.

\begin{figure}[!t]
\centering
\subfloat[Fixed rank]{\includegraphics[width=0.9\columnwidth]{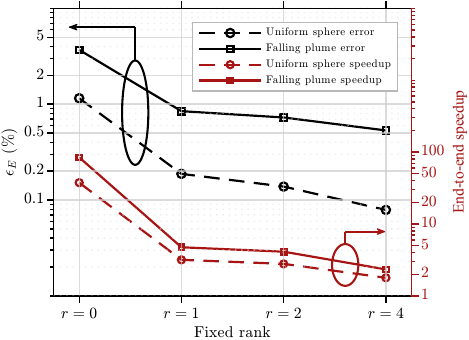}\label{fig:ne1_fixed}}\\[-2pt]
\subfloat[Adaptive rank]{\includegraphics[width=0.9\columnwidth]{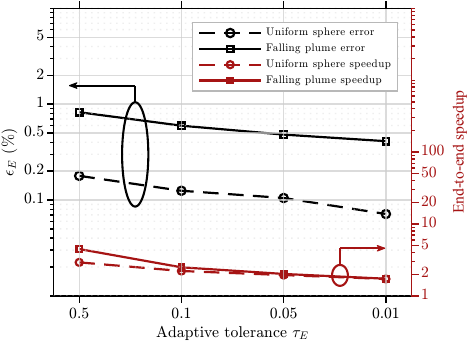}\label{fig:ne1_adaptive}}
\caption{Morphology dependence at matched mean spacing $2\lambda$. Black and red curves denote field error and end-to-end speedup, respectively, with identical axis scales in both panels.}
\label{fig:ne1}
\end{figure}

\subsection{Template-Shape Robustness and Iterative Behavior}
To test shape robustness, the 100,000-chaff uniform-sphere centers and orientations are fixed while only the TWA shape is changed among straight (S), bent (B), and twisted-bent (TB) elements. 
The three inputs therefore share the same realization and S-NFC graph. Fig.~\ref{fig:tb} presents the most complex TB case over the 720-point bistatic cut; the full and rank-one RCS curves are nearly coincident for $\phi_{\rm obs}=0^\circ:0.5^\circ:359.5^\circ$.

For observation direction $\phi_m$, the angular contribution shown by the black curve is
\begin{equation}
\begin{aligned}
\delta_E(\phi_m)
&=\frac{1}{D_E}\bigl[|\widehat E_\theta(\phi_m)-E_\theta^{\rm ref}(\phi_m)|^2\\[-1pt]
&\hspace{18mm}+|\widehat E_\phi(\phi_m)-E_\phi^{\rm ref}(\phi_m)|^2\bigr]^{1/2},\\
D_E
&=\left[\sum_n\left(|E_\theta^{\rm ref}(\phi_n)|^2+|E_\phi^{\rm ref}(\phi_n)|^2\right)\right]^{1/2}.
\end{aligned}
\label{eq:angular_error}
\end{equation}
With this normalization, $\epsilon_E=[\sum_m\delta_E^2(\phi_m)]^{1/2}$; the TB run gives $\epsilon_E=0.188\%$.

\begin{figure}[!t]
\centering
\includegraphics[width=1\columnwidth]{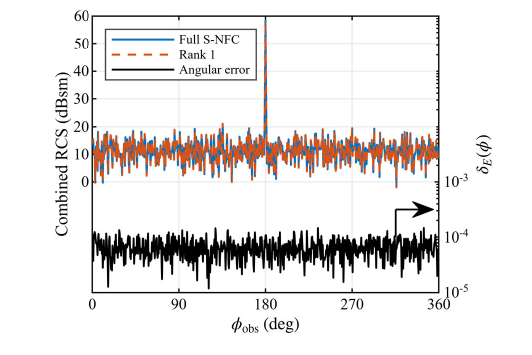}
\caption{TB-cloud 720-point bistatic validation for the 100,000-chaff shape-robustness test. Left axis: combined RCS for full S-NFC and rank one. Right logarithmic axis: globally normalized angular complex-field discrepancy.}
\label{fig:tb}
\end{figure}

Fig.~\ref{fig:convergence} reports the one-million-chaff straight-cloud residual histories at $f=8.427$~GHz. Ranks one and two reach $10^{-4}$ in 10 iterations, rank four in 15, and full S-NFC in 56. Stronger compression therefore reduces both matrix--vector multiplication cost and GCR iterations. Since ranks one and two remain below $1\%$ error for both sparse $2\lambda$ clouds in Fig.~\ref{fig:ne1}, low ranks reduce per-iteration and iterative work while preserving the selected far-field observable. These counts refer to the respective approximate operators.

\begin{figure}[!t]
\centering
\includegraphics[width=0.91\columnwidth]{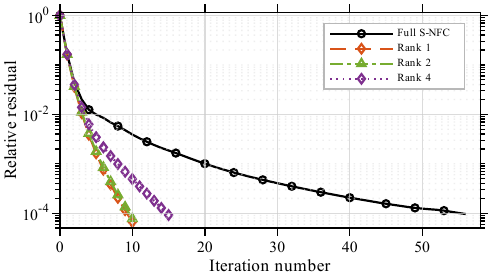}
\caption{GCR residual histories for the one-million-chaff straight-cloud baseline at $f=8.427$~GHz.}
\label{fig:convergence}
\end{figure}

\subsection{Million-Chaff Validation}

Two straight-chaff clouds with $N_c=10^6$ are considered: M1 denotes the uniform cloud, whereas M2 denotes the plume cloud.
They use the same seed and are matched at a measured mean nearest-neighbor spacing of $2.5\lambda$; their mean neighbor counts within $3\lambda$ are 1.236 and 3.030, and their directed retained-block counts are 1,236,058 and 3,029,936.
Thus, the plume has a $2.45\times$ larger retained graph.
The M1/M2 full sweeps take 1.405/5.707~h, reach peak memory of 19.81/38.57~GiB, and average 9.44/25.84 GCR iterations.

Fig.~\ref{fig:million} overlays adaptive error and end-to-end speedup. For M1, $\tauE=0.5$ gives $0.043\%$ error, mean rank 1.000, $8.28\%$ retained-coupling storage, and $2.05\times$ speedup; at $\tauE=0.01$, these values are $0.018\%$, 1.404, $12.61\%$, and $1.73\times$. For M2, the corresponding loose/tight results are $0.253\%/0.117\%$ error, mean ranks $1.001/1.716$, storage ratios $6.60\%/12.42\%$, and speedups $6.92\times/2.65\times$. The greater plume gain is consistent with its larger retained graph and heavier full-system iterative workload.

\begin{figure}[!t]
\centering
\includegraphics[width=0.9\columnwidth]{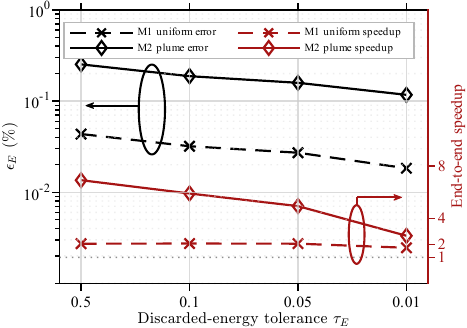}
\caption{One-million-chaff adaptive tradeoff for the uniform cloud (M1) and plume cloud (M2). Black curves use the left logarithmic field-error axis; red curves use the right end-to-end-speedup axis.}
\label{fig:million}
\end{figure}

\FloatBarrier
\section{Conclusion}

Concatenated-row low-rank compression further accelerates S-NFC by sharing a receiving-side basis across retained neighboring interactions while preserving exact self interactions. The required rank depends on local cloud morphology and coupling density, while stronger compression reduces both matrix--vector multiplication cost and generalized conjugate residual iterations. For a one-million-chaff plume, the proposed method stores only $6.60$\% of the retained coupling and achieves a $6.92\times$ end-to-end speedup over uncompressed S-NFC with $0.253$\% complex-far-field error. These results demonstrate the scalability of low-rank-accelerated S-NFC for large-scale full-wave chaff analysis.

\FloatBarrier
\clearpage
\bibliographystyle{IEEEtran}
\bibliography{references}

\end{document}